\documentclass[aps,prb,preprint,superscriptaddress]{revtex4}
\usepackage{graphicx}% Include figure files
\usepackage{dcolumn}% Align table columns on decimal point
\usepackage{bm}% bold math
\usepackage[usenames]{color}
\usepackage{amsmath}

\usepackage{amssymb}

\begin{document}

\preprint{preprint}

\title{Topological stability of vortex configuration in a proximity effect-induced superconducting network}

\author{S. Tsuchiya}
\affiliation{Department of Applied Physics, Hokkaido University,
Sapporo 060-8628, Japan}

\author{S. Tanda}
\affiliation{Department of Applied Physics, Hokkaido University,
Sapporo 060-8628, Japan}
\affiliation{Center of Education and Research for Topological Science and Technology, Hokkaido University,
Sapporo 060-8628, Japan}

\date{\today}% It is always \today, today,

\begin{abstract}
We examine possible effects of the line width of the network for vortex configuration by using Pb-Au bilayer honeycomb network. From the results of Little-Parks oscillation, the superconducting current is found to flow through the path enclosed by edge of wire, not center of wire of the network. Furthermore, in the power spectrum analysis, stable vortex configurations are found to appear in the same way as the case of the Pb monolayer network. Thus the network of arranged vortices, which is based on the dual network, is topologically stable because it is kept regardless of the line width of the base network.
\end{abstract}

\maketitle

\section{Introduction}
Graph or network is a topological concept that represents connectivity with nodes and branches. This concept has been applied to various fields such as an electrical circuit, a molecular structure, and a computer network. Their properties are well affected by the connectivity of network. There are no information in terms of length, width, and curve of branches.\\
\indent Superconducting networks, which consist of multiply connected thin wire of superconductor, are typical examples of them. In superconducting networks, the network structure well affect to the physical properties because this system is sensitive to phase coherence of the order parameter over the network. In fact, phase interference phenomena are driven by the magnetic field known as Little-Parks oscillation. \cite{lpo} Characteristic vortex configurations are caused for various geometries of the network.\cite{tho1,tho2,tho3,exp1,exp2,exp3,exp4,exp5,tho4,tho5,tho6,exp6} So one can observe the network effect as dips or cusps of variation in magnetic field responses. When the line width of the network becomes thick, topology of the network is kept. However it has remained unclear in detail whether vortex configuration is kept or not. In these systems, the line width $d$ is assumed as much smaller than the lattice constant $l$. Moreover, it seems to be difficult to investigate and compare effects of the line width experimentally because measured temperature is close to superconducting transition temperature $T_c$ in most cases.\\
\indent In this paper, to investigate the size effects of the wire for vortex configuration, we propose a superconductor-normal metal bilayer wire network as a new network. When a superconductor and a normal metal are joined with good electrical contact, superconductivity is weakened and induces into the normal metal known as the proximity effect.\cite{degennes} In this case, since the order parameter could be modified along the width direction, novel network effects and size effects are expected to occur. In addition, a relation with studies of superconducting films with antidot arrays \cite{anti1,anti2,anti3,anti4} may be revealed in terms of vortex configuration. Generally, a duality relation is existed between the antidot arrays that $d$ is considered as comparable to $l$ and the superconducting networks. From our results of Little-Parks oscillation, a superconducting current is found to flow through a path enclosed by edge of the wire, not center of the wire and stable vortex configurations are found to appear in the same way as the case of a monolayer wire network. It means that the dual network consisted of vortices is kept regardless of the line width of the base network.

\section{Experimental}
\begin{figure}[h]
\begin{center}
\begin{minipage}{21pc}
\includegraphics[width=0.9\linewidth]{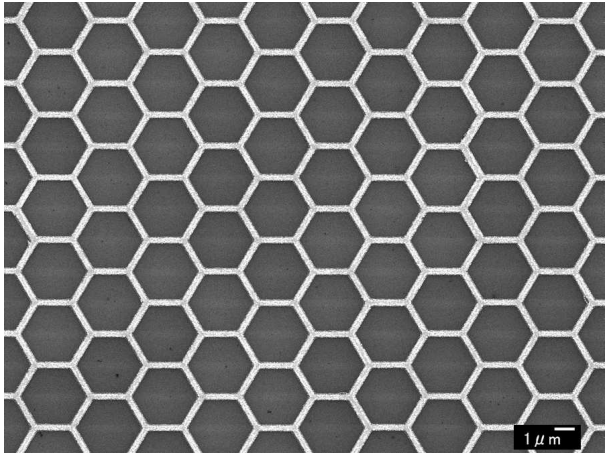}
\end{minipage}
\caption{SEM image of the honeycomb network which has about $2500$ cells with lattice constant of $2$ $\mu$m, line width of $0.2$ $\mu$m.}
\label{fig1}
\end{center}
\end{figure}

To make a comparison, two types of a honeycomb network that we used were fabricated by standard electron beam lithography. One consists of Pb monolayer wire, the other Pb-Au bilayer wire. The gold layer of $0.01$ $\mu$m and the lead layer of $0.1$ $\mu$m are thermally evaporated on a SiO$_2$ substrate followed by the resist lift-off. Figure \ref{fig1} shows a scanning electron microscope (SEM) image of the sample. These samples have about $2500$ cells with lattice constant of $2$ $\mu$m, line width of $0.2$ $\mu$m.\\
\indent Little-Parks oscillation is a powerful tool to investigate the configuration of vortices on the network. Little-Parks oscillation is a periodic variation of $T_c$ with a magnetic field by the superconducting fluxoid quantization. \cite{lpo} Especially when temperature is near $T_c$, phase coherence of the order parameter is stretched over the whole system. Hence variation of $T_c$ is affected by vortex configuration. Experimentally Little-Parks oscillation of $T_c$ can be observed as a periodic variation of resistance with a magnetic field at fixed temperature, which was taken near the midpoint of normal-to-superconducting transition. We measured Little-Parks oscillation of these samples by using $12.5$ Hz four terminal ac Resistance Bridge with excitation voltage $10$ $\mu$V. Moreover the spectral analysis was performed by maximum entropy method (MEM) to explore other periods.

\section{Results and Discussion}
\begin{figure}[h]
\begin{center}
\includegraphics[width=0.9\linewidth]{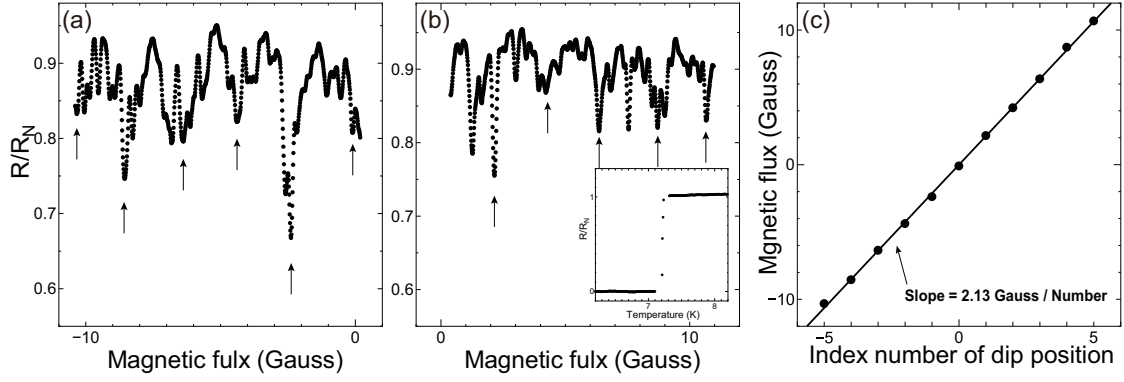}
\caption{(a),(b) The magnetic flux dependence of the sample resistance normalized by $R_N$ in the case of the monolayer network. The arrows indicate periodic dips. The inset shows temperature dependence of the normalized resistance. $T_c$ was observed at around 7.2 K. (c) The index number of dip positions as a function of the magnetic flux. The slope of the line is  $2.13$ G.}
\label{fig2}
\end{center}
\end{figure}

\begin{figure}[h]
\begin{center}
\includegraphics[width=0.9\linewidth]{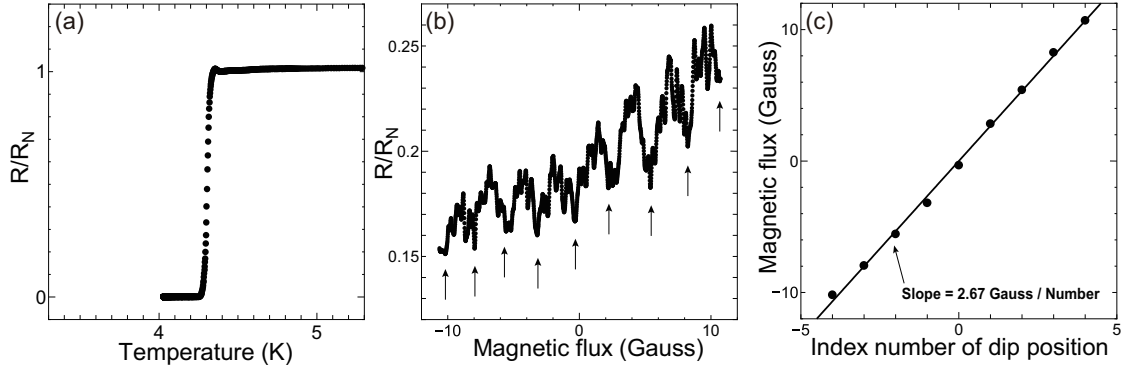}
\caption{(a) Temperature dependence of the sample resistance normalized by $R_N$ in the case of the bilayer network. $T_c$ was observed at around 4.3 K less than 7.2 K. (b) The magnetic flux dependence of the sample resistance normalized by $R_N$ in the case of the bilayer network. The arrows indicate periodic dips. (c) The index number of dip positions as a function of the magnetic flux. The slope of the line is  $2.67$ G.}
\label{fig3}
\end{center}
\end{figure}

\begin{figure}[h]
\begin{center}
\includegraphics[width=0.9\linewidth]{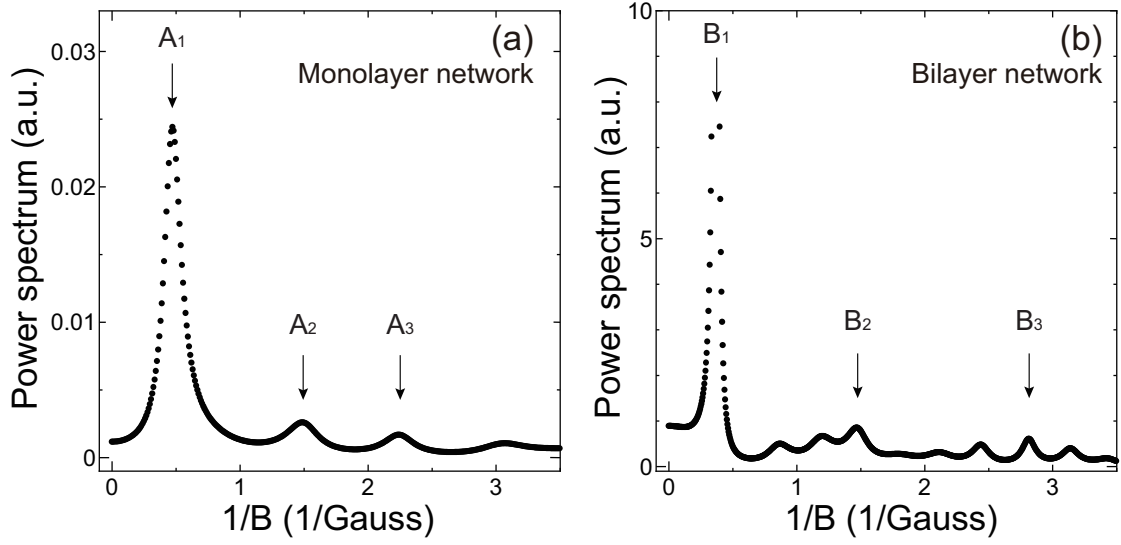}
\caption{(a) The power spectrum in the case of the monolayer network. Fundamental peak labeled as $A_1$ correspond to period of 2.11 G. The second strongest peak labeled as $A_2$ and the third as $A_3$ correspond to a period of $0.67$ and $0.45$ G, respectively. (b) The power spectrum in the case of the bilayer network. Fundamental peak labeled as $B_1$ correspond to period of 2.71 G. The second strongest peak labeled as $B_2$ and the third as $B_3$ correspond to a period of $0.68$ and $0.36$ G, respectively.}
\label{fig4}
\end{center}
\end{figure}

\begin{figure}[h]
\begin{center}
\includegraphics[width=0.8\linewidth]{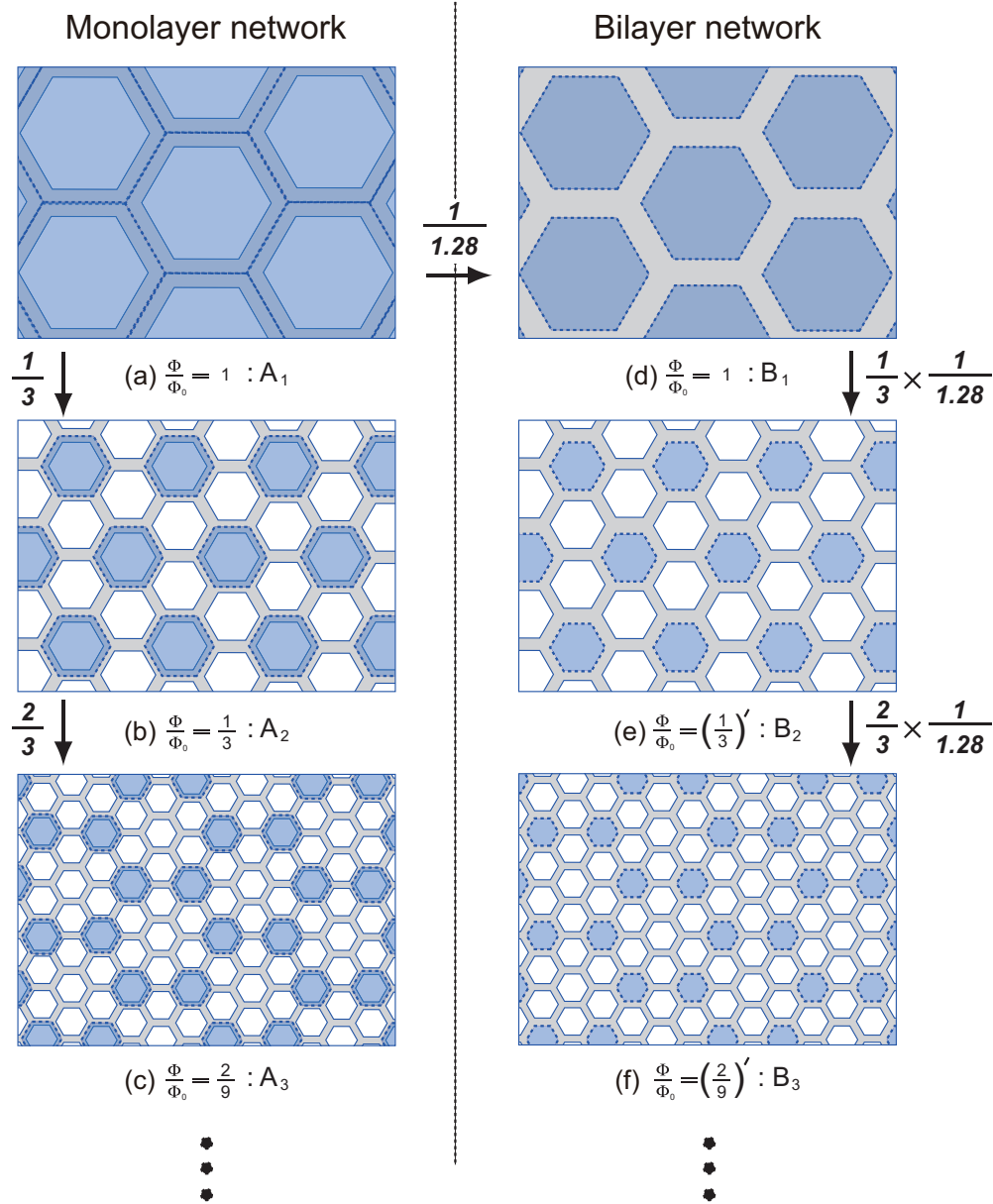}
\caption{(a)-(c) Vortex configuration in the case of the monolayer network at the magnetic flux corresponding to $A_1$, $A_2$, and $A_3$ peak, respectively. The plaquettes occupied by vortices are shown shaded. The dashed line denotes loop of vortex. Vortex consists of loop enclosed by center of the wire. (d)-(f) Vortex configuration in the case of the bilayer network at the magnetic flux corresponding to $B_1$, $B_2$, and $B_3$ peak, respectively. Vortex consists of loop enclosed by edge of the wire. The arrows and the accompanying values indicate the area ratios.}
\label{fig5}
\end{center}
\end{figure}

First the monolayer network was investigated. The inset of Fig. \ref{fig2}(b) present temperature dependence of the sample resistance normalized by $R_N$. $T_c$ was observed at around 7.2 K. This value good agrees with $T_c$ of bulk lead. Figure \ref{fig2}(a) and \ref{fig2}(b) shows the magnetic flux dependence of the normalized resistance in range from $-10$ to $0$ and $0$ to $10$ G. We found periodic dips indicated by the arrows. Figure \ref{fig2}(c) exhibits the index number of dip positions as a function of the magnetic flux. The slope shows period of oscillation as $2.13$ G. The area estimated from the period is $9.72$ $\mu$m$^2$ and correspond to a hexagonal unit cell enclosed by center of the wire of the network. This value compares well to the value $9.85$ $\mu$m$^2$ obtained from SEM observation with $1.3$ $\%$ accuracy. Thus the period correspond to one-flux quantum $\Phi_0 = \hbar /2e$ per unit cell.\\
\indent On the other hand, a different period of oscillation was obtained in the case of the bilayer network. Figure \ref{fig3}(a) shows temperature dependence of the normalized resistance. $T_c$ was observed at around 4.3 K less than 7.2 K. This reduction of $T_c$ demonstrates that the proximity effect is well induced in the bilayer network. \cite{one} Figure \ref{fig3}(b) denotes the magnetic flux dependence of the normalized resistance. Periodic dips indicated by the arrows were found. The slope of the line is $2.67$ G as shown in Fig. \ref{fig3}(c). The area estimated from the period is $7.75$ $\mu$m$^2$ and correspond to a hexagonal unit cell enclosed by edge of the wire, not center of the wire of the network. This value compares well to the value $7.56$ $\mu$m$^2$ obtained from SEM observation with $2.4$ $\%$ accuracy. Hence the area of vortex is found to be different from the case of the monolayer network although both the networks have the same honeycomb design. In this case, the amplitude of the order parameter could be changed along the width direction because the proximity effect modifies the superconducting parameter of the Pb-Au bilayer in terms of the coherence length, the penetration depth, and the extrapolation length.\cite{two} Moreover the normal component of the carriers could be more effective because gold as the normal metal was joined with the superconductor. In addition, superconductivity is weakened in the edge of the wire.\\ 
\indent Spectrum analysis was performed by MEM in order to explore the network effects. Figure \ref{fig4}(a) presents the power spectrum in the case of the monolayer network and total results are summarized in Table \ref{table1} and \ref{table2}. The fundamental peak labeled as $A_1$ correspond to a period of 2.11 G and is consistent with the previous result of $2.13$ G. The corresponding area $S_{A_1}$ is $9.81$ $\mu$m$^2$. In addition, the second strongest peak of $0.67$ G labeled as $A_2$ and the third one of $0.45$ G as $A_3$ are observed. The corresponding area of $S_{A_2}$ and $S_{A_3}$ are $30.90$ $\mu$m$^2$ and $46.00$ $\mu$m$^2$, respectively. An area ratio is connected to the filling ratio of vortex $\Phi/\Phi_0$ which is the magnetic flux $\Phi$ in units of the flux quantum $\Phi_0$ per unit cell. $S_{A_2}/ S_{A_1}$ and $S_{A_3}/ S_{A_2}$ are found to be $3.15$ and $1.49$, respectively. Especially, since $S_{A_2}/ S_{A_1}$ is close to 3, the period of $A_2$ corresponds to the downward cusp in the magnetic field response at $\Phi/\Phi_0 = 1/3$ in the recent report. \cite{rhc} Thus these peaks are attributed to the effect of the honeycomb network.\\
\indent In the case of the bilayer network, several peaks were also observed as shown in Fig. \ref{fig4}(b). The fundamental peak as $B_1$ correspond to a period of 2.71 G and is consistent with the previous result of $2.67$ G. The corresponding area $S_{B_1}$ is $7.64$ $\mu$m$^2$. The second strongest peak labeled as $B_2$ and the third one as $B_3$ correspond to a period of $0.68$ and $0.36$ G, respectively. The corresponding areas of $S_{A_2}$ and $S_{A_3}$ are $30.44$ $\mu$m$^2$ and $57.50$ $\mu$m$^2$. The ratio $S_{B_2}/S_{B_1}$ and $S_{B_3}/S_{B_2}$ are found to be $3.98$ and $1.89$, respectively. These ratios are obviously different from the case of the monolayer network. Therefore these results indicate realization of a novel network attributed to the proximity effect.\\
\indent Now let us discuss vortex configuration on these networks at each peak. In the case of the monolayer network, vortices are commensurately arranged with its base structure at rational $\Phi/\Phi_0$. The ratio $S_{A_2}/S_{A_1}$ and $S_{A_3}/S_{A_2}$ are geometrically expected as 3 and 3/2 because of the honeycomb network. Therefore vortex configuration at $\Phi/\Phi_0 = 1$ corresponding to $A_1$ peak, $\Phi/\Phi_0 = 1/3$ corresponding to $A_2$ peak, and $\Phi/\Phi_0 = 2/9$ corresponding to $A_3$ peak are constructed as illustrated in Fig. \ref{fig5}(a), \ref{fig5}(b), and \ref{fig5}(c), respectively. The unit cells occupied by vortices are shown shaded. The dashed line denotes path through which the superconducting current flows. At $A_1$ peak, vortices are placed in all unit cells enclosed by center of the wire. At $A_2$ peak, one vortex is allocated in every three unit cells. This pattern is commensurately matched with the underlying honeycomb lattice and energetically stable over the whole system. In the same way, stable vortex configuration is constructed at $A_3$ peak. This pattern could be attributed to an effect of the sample edge to keep hexagonal symmetry in addition to the network effect since our network was finite.\\
\indent On the other hand, in the case of the bilayer network, vortex configuration is not simply constructed. At $B_1$ peak, vortices are placed in all unit cells enclosed by edge of the wire as shown in Fig. \ref{fig5}(d). For $B_2$ and $B_3$ peak, their vortex configuration is not constructed soon. The area ratios of $S_{B_2}/S_{B_1}$ and $S_{B_3}/S_{B_2}$ are geometrically expected as 3 and 3/2 in the same way with the case of the monolayer network. But the experimental values are far from the geometrically expected values.\\
\indent Surprisingly, we found a factor associated with the line width in the ratio of $S_{B_2}/S_{B_1}$ and $S_{B_3}/S_{B_2}$ [see Table \ref{table2}]. The ratio of $S_{A_1}/ S_{B_1}$, which means the ratio of area enclosed by edge of the wire and center of the wire, is around $1.28$. Then $S_{B_2}/S_{B_1}$is equal to $3.11\times 1.28$. The 3.11 is very close to $S_{A_2}/ S_{A_1}$ of 3.15. Furthermore $S_{B_3}/S_{B_2}$ is calculated as $1.48\times 1.28$. The 1.48 nearly equal to $S_{A_3}/ S_{A_2}$ of 1.49. That is, the area ratio of the bilayer network is found to consist of two parts. One is the area ratio of the monolayer network, the other the factor of the line width. These relations suggest that the feature of the monolayer network could appear even in the bilayer network. Therefore vortex configuration at the magnetic flux corresponding to $B_2$ and $B_3$ peak are speculated as shown in Fig. \ref{fig5}(c) and (d), respectively. Vortices are arranged in an analogous way with the monolayer network and the path of superconducting current is enclosed by edge of the wire. Vortices seem to be isolated each other. But they are correlated in the same way as the monolayer network.\\
\indent According to our model, topological stability of vortex configuration is revealed. Vortices are correlated each other and networked on a basis of a triangular lattice that is dual to the honeycomb lattice. Thus the network of arranged vortices is topological stable because it is kept regardless of the line width of the base network. This topological stability of vortex configuration could be applied to the antidot arrays system. Furthermore effects of the line width might be studied by controlling the proximity effect.

\section{Summary}
We examine possible effects of the line width of the network for vortex configuration by using Pb-Au bilayer honeycomb network. From the results of Little-Parks oscillation, vortex is found to consist of loop enclosed by edge of the wire, not center of the wire. Furthermore we observed not only fundamental peak but also the second and the third strongest peak in the power spectrum analysis. In comparison with the Pb monolayer network, stable vortex configurations are appeared at each peak in the same way as the case of the monolayer network. Thus the network of arranged vortices, which is based on the dual network, is kept regardless of the line width of the base network. This topological stability of vortex configuration could be applied to the antidot arrays system. Furthermore effects of the line width might be studied by controlling the proximity effect.

\begin{acknowledgements}
The authors are grateful to K. Inagaki, Y. Asano, and S. Uji for useful discussions. We also thank S. Takayanagi, K. Yamaya, and N. Matsunaga for experimental support. This work was supported by Grant-in-Aid for the 21st Century COE program "Topological Science and Technology" from the Ministry of Education, Culture, Sport, Science and Technology of Japan. 
\end{acknowledgements}

\newpage

\begin{table}[htbp]
\begin{ruledtabular}
\caption{The period and the corresponding area.}
\begin{center}
\begin{tabular}{ccccccc}
&$A_1$&$A_2$&$A_3$&$B_1$&$B_2$&$B_3$ \\ \hline
Period (G)&2.11&0.67&0.45&2.71&0.68&0.36 \\
%&$S_{A_1}$&$S_{A_2}$&$S_{A_3}$&$S_{B_1}$&$S_{B_2}$&$S_{B_3}$ \\ \hline
Area ($\mu m^2$)&9.81&30.90&46.00&7.64&30.44&57.50\\
\end{tabular}
\label{table1}
\end{center} 
\end{ruledtabular}
\end{table}

\begin{table}[htbp]
\begin{ruledtabular}
\caption{The area ratio of experimental value, geometrically-expected value, and our findings.}
\begin{center}
\begin{tabular}{cccccc}
&$S_{A_2}/S_{A_1}$&$S_{A_3}/S_{A_2}$&$S_{B_2}/S_{B_1}$&$S_{B_3}/S_{B_2}$&$S_{A_1}/S_{B_1}$\\ \hline
Experimental value&3.15&1.49&3.98&1.89&1.28 \\
Geometrically-expected value&3&$\frac{3}{2}$&3&$\frac{3}{2}$\\
Our findings&&&$3.11\times 1.28$&$1.48\times 1.28$&
\end{tabular}
\label{table2}
\end{center} 
\end{ruledtabular}
\end{table}

\newpage

\newpage

\newpage

\newpage

\end{document}